# Towards a better understanding on agglomeration mechanisms and thermodynamic properties of TiO$_2$ nanoparticles interacting with natural organic matter

**Frédéric Loosli** [a], **Letícia Vitorazi** [b], **Jean-François Berret** [b] and **Serge Stoll** [a,*]

**Affiliations:**
[a] Group of Environmental Physical Chemistry, University of Geneva, F.-A. Forel Institute Section des Sciences de la Terre et de l'Environnement, 10 route de Suisse, 1290 Versoix, Switzerland
[b] Laboratoire Matière et Systèmes Complexes, UMR 7057 Université Paris-Diderot/CNRS, Bâtiment Condorcet, 10 rue Alice Domon et Léonie Duquet, F-75205 Paris cedex 13, France

[*]**Corresponding Author and Address:**
Serge Stoll [*]: Group of Environmental Physical Chemistry, University of Geneva, F.-A. Forel Institute, Section des Sciences de la Terre et de l'Environnement, 10 route de Suisse, 1290 Versoix, Switzerland
Phone: + 41 22 379 0333; Fax: + 41 22 379 0302
email: serge.stoll@unige.ch

## Abstract

Interaction between engineered nanoparticles and natural organic matter is investigated by measuring the exchanged heat during binding process with isothermal titration calorimetry. TiO$_2$ anatase nanoparticles and alginate are used as engineered nanoparticles and natural organic matter to get an insight into the thermodynamic association properties and mechanisms of adsorption and agglomeration. Changes of enthalpy, entropy and total free energy, reaction stoichiometry and affinity binding constant are determined or calculated at a pH value where the TiO$_2$ nanoparticles surface charge is positive and the alginate exhibits a negative structural charge. Our results indicate that strong TiO$_2$-alginate interactions are essentially entropy driven and enthalpically favorable with exothermic binding reactions. The reaction stoichiometry and entropy gain are also found dependent on the mixing order. Finally correlation is established between the binding enthalpy, the reaction stoichiometry and the zeta potential values determined by electrophoretic mobility measurements. From these results two types of agglomeration mechanisms are proposed depending on the mixing order. Addition of alginate in TiO$_2$ dispersions is found to form agglomerates due to polymer bridging whereas addition of TiO$_2$ in alginate promotes a more individually coating of the nanoparticles.

# 1. Introduction

Nanoparticles are produced in increasing quantities due to their unique surface properties (Auffan et al. 2009, Gottschalk et al. 2009) and are used in many domains and applications (Chen and Mao 2007, Lu et al. 2007, Luo et al. 2006). A non negligible amount of these engineered nanoparticles (ENPs) is thus entering aquatic systems as individual or agglomerated nanoparticles through industrial discharges, surface runoff from soils or from wastewater treatment effluent (Batley et al. 2013, Seitz et al. 2012). Once in aquatic environments, interactions with natural compounds will modify their stability, fate, bioavailability and toxic effects towards living organisms (Christian et al. 2008, Handy et al. 2008, Klaine et al. 2008, von Moos and Slaveykova 2014).

Among ENPs, $TiO_2$ nanoparticles are produced in high tonnage (Piccinno et al. 2012) and are present in many customer products such as food, cosmetic and painting industries (Chen and Mao 2007, Cozzoli et al. 2003, Gratzel 2004, Jimin et al. 2010, Khan and Dhayal 2008, Mahltig et al. 2005, Weir et al. 2012, Wongkalasin et al. 2011, Zhang and Sun 2004). $TiO_2$ ENPs production exceeds tons per year (Hendren et al. 2011, Piccinno et al. 2012, Schmid and Riediker 2008) and the expected concentration present in aquatic systems is in the ng $L^{-1}$ to $\mu g\ L^{-1}$ range (Batley et al. 2013, Gottschalk et al. 2009, Sani-Kast et al. 2015). Therefore $TiO_2$ is often considered to study the possible transformations of nanoparticles in aquatic systems. Most of the studies on $TiO_2$ ENPs transformation processes in presence of natural organic matter (NOM) demonstrate that humic substances and non humic substances such as extracellular polymeric substances are found to deeply modify the $TiO_2$ stability (Belen Romanello and Fidalgo de Cortalezzi 2013, Chae et al. 2012, Domingos et al. 2009, Erhayem and Sohn 2014, Gallego-Urrea et al. 2014, Shen et al. 2015, Zhang et al. 2013). Indeed recent findings considering the effect of NOM have shown that electrostatic repulsions and steric interactions between NOM molecules adsorbed onto the ENPs surface enhance the stability and thus the ENPs dispersion state (Hyung et al. 2007, Liu et al. 2010, Loosli et al. 2013, Louie et al. 2013, Palomino and Stoll 2013, Zhang et al. 2009). Moreover NOM complexation with ENPs is also found to induce the partial redispersion of already formed ENP agglomerates (Baalousha 2009, Loosli et al. 2014, Mohd Omar et al. 2014). All these studies are investigating the ENPs surface charge modification and give a special interest to the ENPs state of dispersion (resulting size, charge and stability of the ENPs) in presence of NOM. In the present study, a particular concern is given to the quantification of thermodynamic interaction parameters using isothermal titration calorimetry (ITC) for the investigation of interactions between anatase $TiO_2$ ENPs and alginate, an extracellular polymeric substance model (Bernhardt et al. 1985, Gregor et al. 1996). Indeed polysaccharides represent 10-30% of the NOM in lakes (Buffle et al. 1998, Wilkinson et al. 1999, Wilkinson et al. 1997) whereas in marine environment polysaccharide content may be as high as 50% of the dissolved organic carbon (Engel et al. 2004, Wells 1998). Alginate is a naturally occurring polysaccharide and is found in aquatic environments (Gombotz and Wee 1998). It is produced by brown algae species but also by some bacteria (Gombotz and Wee 1998, Saude et al. 2002). Alginate is a negatively charged linear block copolymer composing of 1→ 4 linked β-D-mannuronic acid and its C-5 epimer α-L-guluronic acid. Alginate is also used as a thickener agent in food industry and as drug carrier in biomedicine (Helgerud 2009, Lee and Mooney 2012). ITC is a universal method which is suitable to follow the energies of association reactions. Indeed important parameters that are global properties of the systems such as the change of enthalpy ΔH, the change of entropy ΔS and the change of the total free energy ΔG that occur during the binding processes can be extracted. ITC measurements also provide information on the interaction affinity by determination of the affinity binding constant $K_b$ and the interaction reaction stoichiometry. All these parameters are accessible through a single experiment. This method has been applied to a wide range of chemical and biochemical binding interactions and was especially used for protein-substrate interactions, structure-based drug design and supramolecular polymers self-associations (Arnaud and Bouteiller 2004, Brinatti et al. 2014, Cedervall et al. 2007, Chiappisi et al. 2013, Doyle 1997, Jelesarov and Bosshard 1999, Kamiya et al. 1996, Kim et al. 2010, Ladbury 2001, Matulis et al. 2002, Vitorazi et al. 2014). Humic acid coverage on arsenic was shown to slightly reduce the binding interaction and the rate constant between arsenic-coated and ferrihydrite-kaolinite mixtures (Martin et al. 2009). Sheng et al. investigated the binding properties of copper ions with extracellular polymeric substances and showed that the process was mainly driven by entropy (Sheng et al. 2013). Another study on alginate-sodium dodecyl sulfate interaction shown that aggregation was due to hydrophobic interactions (Yang et al. 2008). Thermodynamic adsorption profile at a solvated organic-inorganic interface was done and gold ENPs interaction with carboxylic acid-terminated alkanethiols was found exothermic and enthalpy driven (Joshi et al. 2004, Ravi et al. 2013).

The present work is dealing with a novel approach that has a high potential to contribute to a better understanding of the behavior of ENPs in the presence of NOM. A detailed description is achieved on the complex interaction phenomena which occur between ENPs and NOM by determining key thermodynamic parameters and their link with ENPs surface charge properties to understand agglomeration mechanisms. Such quantitative information is often missing when ENP interactions with aquagenic compounds have to be investigated.

## 2. Materials and methods
### 2.1. Materials
$TiO_2$ engineered nanoparticles were obtained from Nanostructured & Amorphous Material Inc (Houston, TX, USA) as a 15% wt suspension of 15 nm anatase nanoparticles in water (Loosli et al. 2013). This stock solution was then, after homogenization, diluted by adding Milli Q water (Millipore, Zoug, ZG, Switzerland, with R >18 MΩ.cm, T.O.C. <2 ppb) to reach a final 5 g $L^{-1}$ $TiO_2$ mass concentration. For sodium alginate (A2158, Sigma Aldrich, Buchs, SG, Switzerland) a 10 mM solution in term of charge concentration was prepared by dissolving the low viscosity biopolymer in MilliQ water and by stirring it overnight. NaOH and HCl (1 M, Titrisol®, Merck, Zoug, ZG, Switzerland) were used after dilution to adjust the dispersions and solutions at pH 3.1 and 11.0. The compounds were dialyzed simultaneously into the same water buffer using separate 12-14 kDa cutoff dialysis membranes (Spectrum Laboratories, Inc., Rancho Dominguez, CA, USA) to minimize the change of enthalpy (ΔH) from dilution process during titration. The solvent from the dialysis was used to dilute the $TiO_2$ and alginate suspensions to experimental concentrations (from 0.1 to 1.4 g $L^{-1}$ for $TiO_2$ and from 0.05 to 2.5 mM for alginate). Finally, prior utilization they were filtered through a 0.45 µm cellulose acetate filter (VWR, Nyon, VD, Switzerland) to remove possible presence of laboratory air dust. No loss of $TiO_2$ and alginate was observed during the filtration process.

## 2.2. Isothermal titration calorimetry measurement

The heat exchange between $TiO_2$ ENPs and alginate was determined using a VP-ITC calorimeter (MicroCal Inc., Northampton, MA, USA) with a sample cell volume equal to 1.4643 mL. The working temperature was set to 298.15 K and, after a preliminary 2 µL ligand (L) injection, 28 injections of 10 µL of ligand into the sample cell containing 1.4643 mL (which is equal to the cell volume) of the macromolecule (M) were realized with an injection duration of 20 seconds and a 210 seconds delay between each successive injection. The agitation speed was set to 307 rpm. Equation (Eq. (1)) to fit the data representing the heat of exchange ($dQ/dn_L$) during the association process as a function of the molar charge ratio (Z = [L]/[M], where [L] and [M] are the ligand and macromolecule molar charge concentration) was derived from the Multiple Non-Interacting Sites (MNIS) model where binding sites are considered to be independent (Courtois and Berret 2010).

$$\frac{dQ}{dn_L}(Z) = \frac{1}{2} \Delta H_b \left[ 1 + \left(1 - \frac{[L]}{n[M]} - \frac{1}{n K_b [M]}\right) \left(\left(1 + \frac{[L]}{n[M]} + \frac{1}{n K_b [M]}\right)^2 - \frac{4[L]}{n[M]}\right)^{-\frac{1}{2}} \right]$$
(1)

In Eq. (1) the fitting parameters $\Delta H_b$ in kJ $mol^{-1}$, $K_b$ in $M^{-1}$ and n represent the binding enthalpy, binding constant and reaction stoichiometry respectively. The system total free energy, $\Delta G$ in kJ $mol^{-1}$, and entropy, $\Delta S$ in kJ $K^{-1}$ $mol^{-1}$, are calculated from the fitting parameters ($K_b$ and $\Delta H_b$ values) with $\Delta G = -RTlnK_b$ and $\Delta S = (\Delta H - \Delta G)/T$. The ligand and macromolecule charge numbers for the ith-injection are equal to:

$$L_i = L_{i-1} + V_i \cdot [L]_m \cdot N_A \quad and \quad M = [M]_m \cdot V_{cell} \cdot N_A \quad (2)$$
$$with \quad [TiO_2]_m = [TiO_2]_M \cdot S_A \cdot \sigma_{TiO_2} \cdot 10^{18} \div N_A \quad and \quad (3)$$
$$[Alginate]_m = [Alginate]_M \div MW(monomer) \cdot \alpha \quad (4)$$

In Eq. (2), $V_i$ and $V_{cell}$ represent the volume of ligand injected and the cell volume respectively, [L or M]$_m$ correspond to the mol of charge of ligand or macromolecule per unit of volume and $N_A$ to the Avogadro constant. The concentration in term of mol of charge per unit of volume for $TiO_2$ and alginate are expressed in Eq. (3) and (4) respectively. In Eq. (3), $[TiO_2]_M$ represent the $TiO_2$ mass concentration, $S_A$ the ENPs specific surface area and $\sigma_{TiO_2}$ the $TiO_2$ hydroxyl sites density. In Eq. (4) [Alginate]$_M$ represents the alginate mass concentration and α the degree of ionization.

$TiO_2$ charge concentrations were calculated based on manufacturer data of the primary $TiO_2$ diameter (equal to 15 nm and in good agreement with the mode value of the number size distribution determined by dynamic light scattering, Fig. S2b) to calculate $S_A$ and on the other hand on a previous study to obtain $\sigma_{TiO_2}$ (Kominami et al. 2000). A $S_A$ and $\sigma_{TiO_2}$ equal to 100 m$^2$ g$^{-1}$ and 5 sites nm$^{-2}$ respectively were used to determine the TiO$_2$ charge concentration. The factor of conversion between the mass and charge concentration was thus set so that a 1 g L$^{-1}$ TiO$_2$ dispersion corresponds to a 0.83 mM charge concentration. The degree of ionization of alginate was set equal to 1 (completely deprotonated) due to the net polymer carboxylic acid moieties pK$_a$ decrease in presence of strongly oppositely charged particles (Carnal and Stoll 2011).

Furthermore the mixing order of the two compounds was investigated to better understand the TiO$_2$-alginate and alginate-TiO$_2$ interactions and the agglomeration process as illustrated in Fig. 1. In a first set of experiments (type I) alginate was playing the role of ligand and was added to the TiO$_2$ dispersion. In the second set of experiments realized (type II) TiO$_2$ ENPs (L) were added to alginate (M).

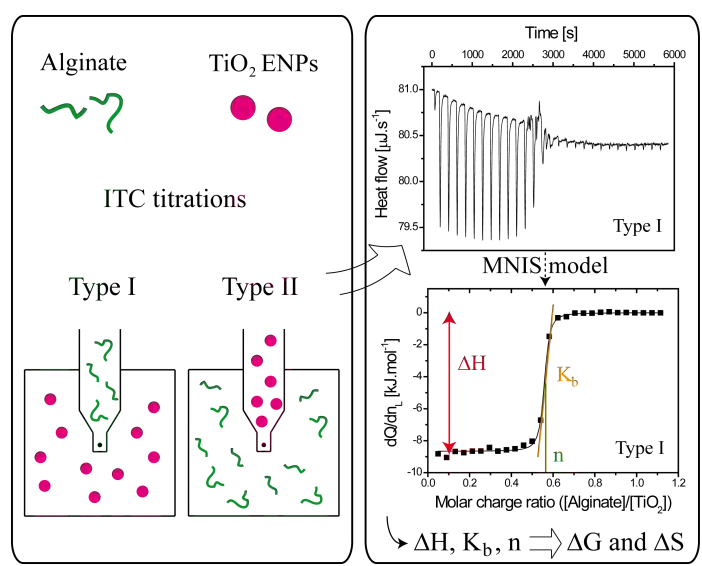

**Fig. 1** - ITC type I (alginate in TiO$_2$ dispersion) and type II (TiO$_2$ in alginate) titrations. ITC measurements give the heat flow for each of the 28 injections and after fitting the integrated data with the multiple non-interacting sites (MNIS) model the enthalpy of exchange ΔH, the binding constant K$_b$ and the reaction stoichiometry n are determined. It permits the calculation of the entropy ΔS and total free energy ΔG for TiO$_2$-alginate interaction.

Experiments were realized at pH 3.1 (and at pH 11.0 for negative controls) without addition of electrolyte. Such a pH value was utilized to address the interaction between isolated, dispersed ENPs and alginate. It is important to note that results presented here remains valid as long as pH < pH$_{PCN,TiO_2}$. Moreover this pH corresponds to the commercial dispersion pH (after dilution to a 5 g L$^{-1}$ concentration) and was chosen to avoid any pH surface charge modification effects. The domain of concentration investigated was from 0.5 mM alginate in 0.1 g L$^{-1}$ TiO$_2$ to 2.5 mM alginate in 0.5 g L$^{-1}$ TiO$_2$ for Type I experiments and from 1 g L$^{-1}$ TiO$_2$ in 0.05 mM alginate to 5 g L$^{-1}$ TiO$_2$ in 0.25 mM alginate for Type II experiments. Such TiO$_2$ concentration are higher than the expected environmental concentration, which are in the ng to μg L$^{-1}$ range (Batley et al. 2013, Gottschalk et al. 2009, Sani-Kast et al. 2015), because of the importance to obtain an optimum signal with the calorimeter. Also it should be noted that concentrations could be significantly higher than the expected environmental concentration in case of local pollution events.

### 2.3. Zeta potential and size distribution measurements

Zeta ($\zeta$) potential values and z-average hydrodynamic diameters of $TiO_2$ dispersion and alginate solution as a function of pH as well as $TiO_2$ in presence of alginate (for Type I and II titrations) as a function of charge ratio were determined by measuring the particles velocity using laser doppler velocimetry with phase analysis light scattering (M3-PALS) and diffusion coefficient using dynamic light scattering (DLS) (Zetasizer Nano ZS instrument, Malvern Instruments, Worcestershire, UK). The instrument was operating at 298.15 K with a 4 mW He-Ne laser working at 633 nm and the detection angle for DLS measurement was 173° (back scattering). For $\zeta$ potential values determination the Smoluchowski approximation model was applied according to the formation of large agglomerates (Baalousha 2009). All polydispersity indexes were found below 0.6.

## 3. Results and discussion

To determine the binding properties between $TiO_2$ ENPs and alginate, ITC experiments were realized at pH < $pH_{PCN,TiO_2}$ where strong electrostatic interactions occur between the negatively charged alginate and positively charged $TiO_2$ ENPs. As shown in Fig. 2, in which the $\zeta$ potential values of $TiO_2$ and alginate are represented as a function of pH, at pH 3.1 (large gray vertical line) the ENPs exhibit a positive surface charge ($\zeta$ potential = +40.9 ± 1.4 mV (mean ± standard deviation on mean of triplicates)) and the biopolymer structural charge is negative with a $\zeta$ potential value found equal to -13.0 ± 0.1 mV. At pH < $pH_{PCN,TiO_2}$ the $TiO_2$ ENPs are dispersed with a z-average diameter equal to 47 ± 1 nm (Figs. S1 and S2 in Supporting information) and alginate z-average diameter is equal to 178 ± 21 nm (Figs. S3 and S4).

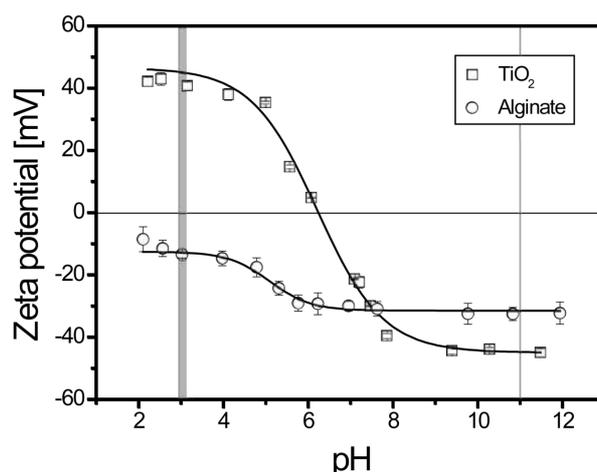

**Fig. 2** - Zeta potential values of $TiO_2$ (open squares) and alginate (open circles) as a function of pH. $TiO_2$ isoelectric point is found here equal to 6.2 ± 0.1 whereas alginate exhibit a negative structural charge in the full pH range. At pH 3.1 $TiO_2$ and alginate have $\zeta$ potential values equal to +40.9 ± 1.4 mV and -13.0 ± 0.1 mV respectively (large gray vertical line). At pH 11.0 both compounds are negatively charged (narrow gray vertical line). $[TiO_2]$ = 50 mg $L^{-1}$, [Alginate] = 100 mg $L^{-1}$ and [NaCl] = 0.001 M.

## 3.1. Determination of TiO$_2$-Alginate thermodynamic binding parameters by ITC

**Titration of TiO$_2$ dispersion with alginate (Type I)** When alginate is added to TiO$_2$ ENPs strong interactions are observed as shown in Fig. 3a which represents a thermogram of the heat of exchange as a function of time for a 100 mg L$^{-1}$ TiO$_2$ dispersion titrated by a 0.5 mM alginate charge concentration. Generally when adding alginate to the TiO$_2$ dispersion the instrument responds so as to compensate the binding reaction exchange energy to maintain a small difference of temperature between the reference cell and the "reaction" cell constant. Both cells are isolated in an adiabatic jacket. The heat compensation is recorded and in the present case "negative" peaks reveal an exothermic process. Each peak corresponds to a consecutive alginate addition. In the thermogram the first response (first peak) is of lower intensity than the following owing to the smaller alginate volume injected (2 $\mu$L instead of the 10 $\mu$L "normal" injection volume). For the next ten injections the peak amplitudes are constant. It means that the number of TiO$_2$ free sites available for random alginate adsorption is high enough to undergo adsorption of a maximum and identical alginate amount. Then the peak intensity rapidly decreases as less positively TiO$_2$ sites are available and finally only low heat exchange corresponding to dilution effect is observed for the last injections due to sites saturation (exchange energy identical to alginate titration in water at pH 3.1, as shown in Fig. S5, with dilution giving rise to low exothermic interaction).

From the data presented in Fig. 3a, the variation of the exchange energy as a function of alginate over TiO$_2$ charge ratio (Z = [Alginate]/[TiO$_2$]) is plotted in Fig. 3b. The values of interaction heat of exchange (dQ/dn$_L$) for each injection are equal to the corresponding peak area integration from the real-time thermogram. For the 100 mg L$^{-1}$ TiO$_2$ titration with 0.5 mM alginate the binding enthalpy is found equal to -8.7 kJ mol$^{-1}$ which indicates an exothermic interaction process. K$_b$ is equal to 3.5 × 10$^7$ M$^{-1}$ which points out a high binding affinity between TiO$_2$ and alginate in this electrostatic scenario (positively charged TiO$_2$ ENPs in the presence of negatively charged alginate). The reaction stoichiometry is found equal to 0.56. All these fitting parameters (derived from Eq. (1)) allow the calculation of the free energy ΔG and entropy ΔS changes during the interaction process. These energy changes are equal to -4.3 × 10$^4$ kJ mol$^{-1}$ and +115.3 J mol$^{-1}$ K$^{-1}$ respectively. Two other experiments with different concentrations but with the same [L]/[M] ratio have been done to evaluate the effect of relative concentration on binding processes. The real-time thermograms and respective integrated heat data fitted with the MNIS model for these experiments are represented in Figs. S6 and S7 and all the fitting and calculated parameters derived from the type I titrations are listed in Table 1.

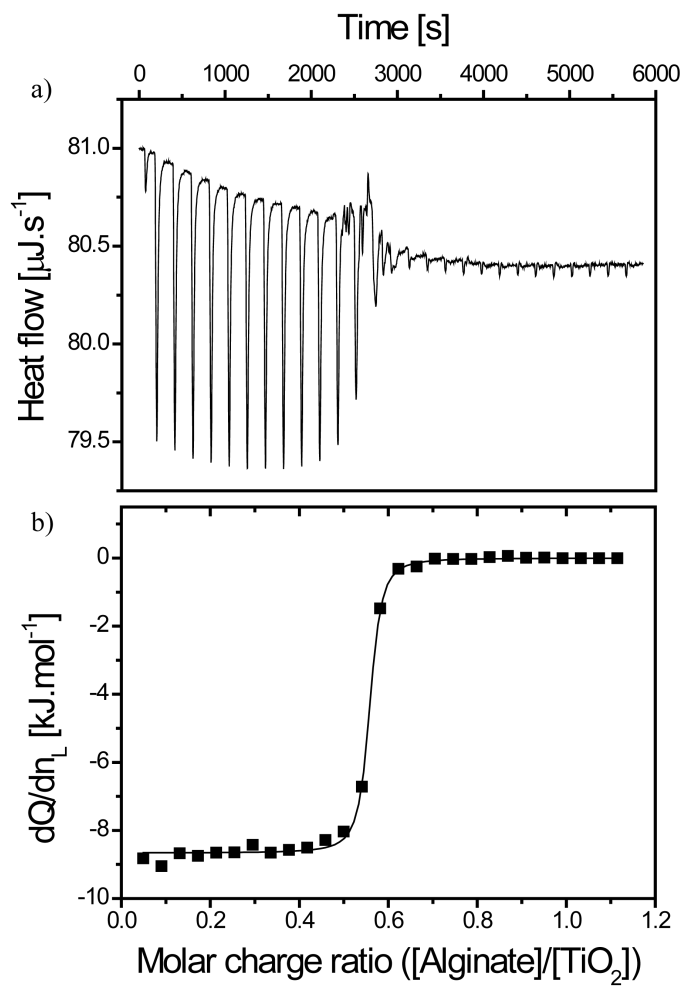

**Fig. 3** - a) Real-time thermogram for $TiO_2$ 0.1 g $L^{-1}$ titration with alginate 0.5 mM at pH < $pH_{PCN,TiO_2}$ and at 298.15 K. The heat flow refers to the thermal compensation of the calorimeter to keep the sample at a constant temperature. Here negative peaks indicate an exothermic reaction. After about fifteen injections sites saturation occurs and only dilution effect is observed (small negative peaks). b) The respective integrated heat data ($dQ/dn_L$) as a function of molar charge ratio ([Alginate]/[$TiO_2$]) is fitted with the multiple non-interacting sites (MNIS) model. The binding enthalpy, the binding constant and the reaction stoichiometry where found here equal to -8.7 kJ $mol^{-1}$, 3.5 × $10^7$ $M^{-1}$ and 0.56, respectively.

**Table 1:** Fitting parameters $\Delta H_b$, $K_b$ and n from ITC analysis of the integrated heats with MNIS model and calculated $\Delta G$ and $\Delta S$ from $K_b$ and $\Delta H_b$ values. The formation of complexes between $TiO_2$ ENPs and alginate is found spontaneous, mainly driven by entropic effects and enthalpically favorable.

| Alginate in $TiO_2$ | $\Delta H_b$ [kJ $mol^{-1}$] | $K_b$ [$M^{-1}$] | n | $\Delta G$ [kJ $mol^{-1}$] | $\Delta S$ [J $K^{-1}$ $mol^{-1}$] |
|---|---|---|---|---|---|
| 0.5 mM in 0.1 g $L^{-1}$ | -8.7 | 3.5 × $10^7$ | 0.56 | -43.0 | 115.3 |
| 0.7 mM in 0.14 g $L^{-1}$ | -9.0 | 2.4 × $10^7$ | 0.56 | -42.2 | 111.3 |
| 2.5 mM in 0.5 g $L^{-1}$ | -8.7 | 1.3 × $10^7$ | 0.61 | -40.6 | 107.2 |

Alginate interaction with $TiO_2$ ENPs is found to be a spontaneous process with high values of Gibbs free adsorption energy ($\Delta G$ <-40 kJ mol$^{-1}$). The binding enthalpy is favorable to the formation of complexes ($\Delta H_b$ <0) and found independent on the compounds concentration. The main interaction driving process is due to an important gain in entropy (-T$\Delta S$ <$\Delta H_b$) with T$\Delta S$ value of about +30 kJ mol$^{-1}$ in our experimental conditions. The binding constant is found to decrease when increasing concentration as $K_b \sim c^{-2}$. It implicates the decrease of the calculated Gibbs free energy and entropy (Courtois and Berret 2010) as the binding enthalpy change is constant and independent on concentration (-8.8 ± 0.2 kJ mol$^{-1}$). The charge stoichiometry was found constant with n = 0.58 ± 0.04.

**Titration of alginate with $TiO_2$ (Type II)** The ITC thermogram and the respective integrated heat data for a $TiO_2$ 1 g L$^{-1}$ and 0.05 mM alginate reaction are shown in Fig. 4. The interaction is also found exothermic and for a charge ratio around Z = 1 no further interaction between $TiO_2$ and alginate occurs and only dilution effect is observed (same exchange heat than in Fig. S8 for $TiO_2$ titration in water). In Table 2 thermodynamic parameters are listed for experiments at different concentrations (Figs. S9 and S10). The binding enthalpy is found independent of the concentration and equal to -8.7 ± 0.4 kJ mol$^{-1}$ and the complexation process is also driven by an important entropy gain.

**Table 2:** Fitting parameters $\Delta H_b$, $K_b$ and n from ITC analysis of the integrated heats with MNIS model and calculated $\Delta G$ and $\Delta S$ from $K_b$ and $\Delta H_b$ values. The formation of complexes between $TiO_2$ ENPs and alginate is found spontaneous, mainly driven by entropic effects and enthalpically favorable.

| $TiO_2$ in Alginate | $\Delta H_b$ [kJ mol$^{-1}$] | $K_b$ [M$^{-1}$] | n | $\Delta G$ [kJ mol$^{-1}$] | $\Delta S$ [J K$^{-1}$ mol$^{-1}$] |
|---|---|---|---|---|---|
| 1 g L$^{-1}$ in 0.05 mM | -8.2 | 1.6 × 10$^7$ | 0.82 | -41.1 | 110.3 |
| 1.4 g L$^{-1}$ in 0.07 mM | -9.0 | 1.4 × 10$^7$ | 0.96 | -40.8 | 106.6 |
| 5 g L$^{-1}$ in 0.25 mM | -8.8 | 7.9 × 10$^6$ | 0.93 | -39.4 | 102.6 |

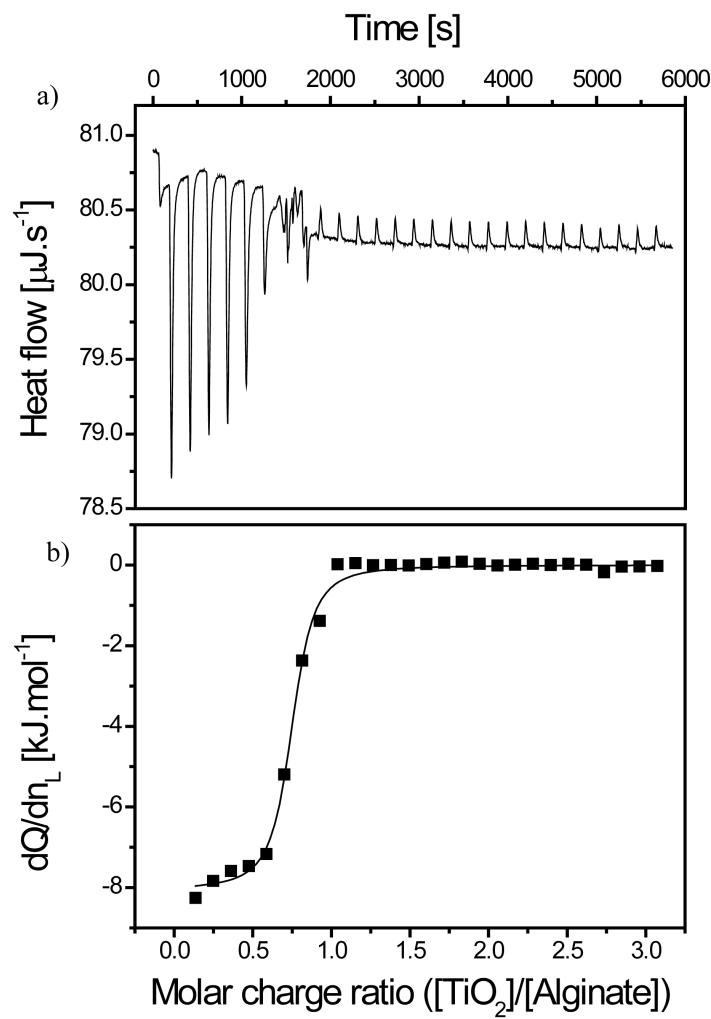

**Fig. 4** - a) Real-time thermogram for alginate 0.05 mM titration with $TiO_2$ 1 g $L^{-1}$ at pH < $pH_{PCN,TiO_2}$ and at 298.15 K. Negative peaks indicate an exothermic reaction and after about ten injections sites saturation occurs and only dilution effect is observed (small positive peaks). b) The respective integrated heat data as a function of $TiO_2$ over alginate molar charge ratio is fitted with the MNIS model. The binding enthalpy, the binding constant and the reaction stoichiometry are found here equal to -8.2 kJ $mol^{-1}$, 1.6 × $10^7$ $M^{-1}$ and 0.82, respectively.

When comparing the two titration types similar binding enthalpy value are observed. Both types of interaction are enthalpically favorable but mainly driven by entropic gain arising from the alginate and ENP counter-ions and water molecules release during adsorption processes. Similar total free energy changes were observed during the association process between ZnO NPs with lysozyme, as well as between proteins and amino acid functionalized gold NPs (Chakraborti et al. 2010, Chatterjee et al. 2010, De et al. 2007). However it should be noted here that the interaction process is mainly driven by an important gain of entropy whereas in the other studies (Chakraborti et al. 2010, Chatterjee et al. 2010) the strong interactions ($K_b$ = 0.9 × $10^6$ $M^{-1}$) were enthalpically favorable but entropically unfavorable due to conformational restriction of proteins. Reactions stoichiometry close to unity clearly indicates that the interaction is electrostatic. When working in conditions where electrostatic interactions are favorable (positively charged ENPs and negatively charged polysaccharides) a stoichiometry close to unity in term of molar charge ratio denotes an electrostatic driven process for compounds in which charges are not sterically hindered (which is the case here for both $TiO_2$ and alginate) and where other functional groups (such as alcohol) are not playing main roles. Nevertheless the stoichiometry of the two titration types is different and explained by different mechanisms of interaction depending on the mixing order. For type I, when alginate is added to $TiO_2$, the polysaccharide is prompt to facilitate bridging between the ENPs which restrict the access to positive $TiO_2$ charges due to conformational hindrance during bridging agglomeration process. Consequently the stoichiometry is smaller than unity (n = 0.58 ± 0.04).

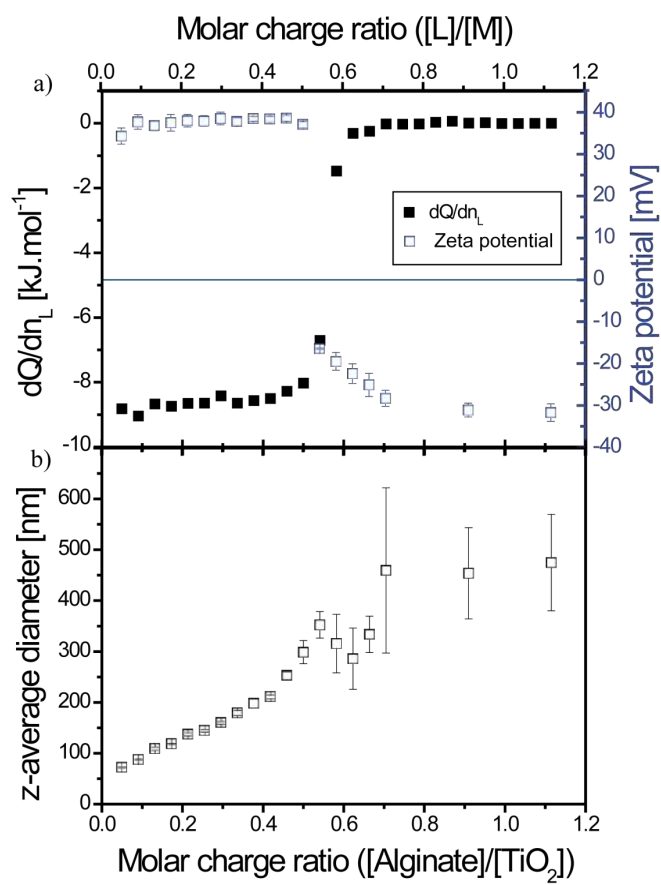

**Fig. 5 -** a) Integrated heat data and $\zeta$ potential values as a function of alginate over $TiO_2$ charge ratio for $TiO_2$ 0.1 g L$^{-1}$ titration with alginate 0.5 mM at pH < pH$_{PCN,TiO_2}$. For a ratio up to 0.5, the binding enthalpy and the $\zeta$ potential values are constant and found equal to - 8.8 ± 0.3 kJ mol$^{-1}$ and +37.6 ± 0.3 mV. Then charge inversion ($\zeta$ potential = -16.4 ± 0.2 mV) is observed for Z = 0.54 and for Z > 0.70 sites saturation occurs and no more interaction is observed. b) z-average diameter as a function of molar charge. Strong $TiO_2$ ENPs destabilization occurs for Z > 0.5 whereas, below this ratio, z-average diameter increase is linear indicating ENPs bridging.

For type II titration, when $TiO_2$ is added to alginate the ENPs are more closely individually coated and the charge stoichiometry is closer to the unity (n = 0.90 ± 0.07). These different mechanisms are related to the initial predominant compound concentration and physico-chemical properties in the reaction cell. The bridging mechanism in type I is favorable to higher entropy gain (ΔS columns in Tables 1 and 2). Indeed, in addition to entropy gain due to the polymer counter-ions and water molecules release, alginate is expected to be less collapsed onto the ENPs surface and thus lower conformational entropy loss is observed than in the type II titration. This conformational entropy gain is nevertheless negligible in comparison to the entropy gain due to counter-ions and water molecule release. Another similarity between the type I and type II titrations is the heat exchange signature on real-time thermogram for the injection just prior binding sites saturation which is the consequence of an important system physical change (as discussed below).

To ensure that electrostatic interactions are governing the complexation process a negative control was done at pH 11.0. At this pH both the $TiO_2$ and the alginate exhibit negative surface and structural charges (Fig. 2), respectively, and no interaction was observed due to electrostatic repulsions between the two compounds (Fig. S11).

### 3.2. Influence of $TiO_2$-alginate complexes structural charge on binding heat of exchange

Determination of the ζ potential and hydrodynamic diameter values for the type I and ζ potential values for type II titrations are done here to evaluate charge modification during titration process and correlate it with the heat exchange values determined from ITC measurements. The time between each successive ligand addition was equal to 210 seconds similarly to the ITC titrations. Hydrodynamic diameters values for type II titration were not determined owing to the insufficient DLS signal until the $8^{th}$ injection (sites saturation).

**Type I titration:** The binding enthalpy and ζ potential values as a function of molar charge ratio for type I titration, alginate 0.5 mM in $TiO_2$ 0.1 g $L^{-1}$ at pH 3.1, are represented in Fig. 5 . For an alginate over $TiO_2$ charge ratio up to 0.50, binding enthalpy per mol of injectant as well as the ζ potential are constant and equal to -8.8 ± 0.3 kJ $mol^{-1}$ and +37.6 ± 0.3 mV respectively. High ζ potential values in this molar charge ratio domain indicates that $TiO_2$ surface coverage is far to be complete and enough $TiO_2$ positively charged binding sites are available for further alginate adsorption which is confirmed by constant and maximum interaction enthalpy values. Then for a ratio of 0.54 $TiO_2$ charge inversion is observed (ζ potential = -16.4 ± 0.2 mV) and the corresponding binding enthalpy is decreasing to -1.6 kJ $mol^{-1}$. For the three next alginate injections the ζ potential values and the binding enthalpy slightly decrease and, for a charge ratio greater than 0.70, sites saturation occurs and no interaction between alginate and $TiO_2$ is recorded, due to electrostatic repulsions and steric effects at high alginate concentration as shown by Lin et al. (Lin et al. 2012), with a corresponding ζ potential plateau value at -30.4 ± 1.8 mV. The z-average diameter of the $TiO_2$ ENPs as a function of the molar charge ratio Z is represented in Fig. 5b. Two domains in term of size variation are observed. In the first one, for a charge ratio below 0.5, the $TiO_2$ z-average diameter increases linearly when alginate is added. This is due to the alginate polymer bridging between the ENPs. Then a second domain is reached for further alginate addition were stronger destabilization of $TiO_2$ is observed due to surface charge neutralization and inversion which result in particle precipitation.

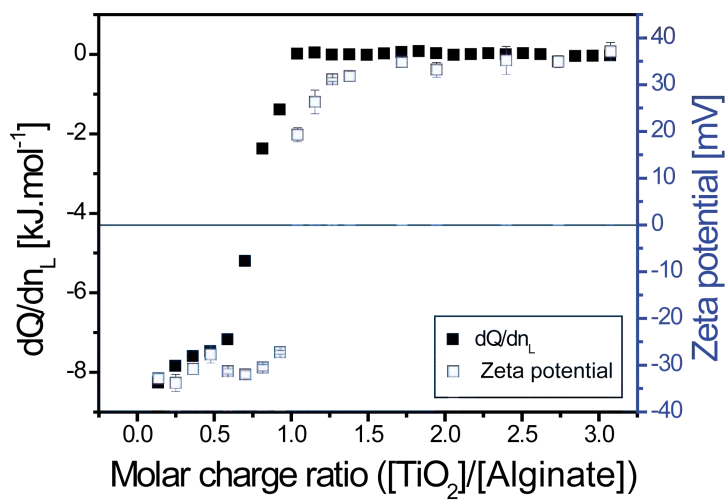

**Fig. 6** - Integrated heat data and $\zeta$ potential values as a function of $TiO_2$ over alginate charge ratio for an alginate 0.05 mM titration with $TiO_2$ 1 g $L^{-1}$ at pH < $pH_{PCN,TiO_2}$. For a charge ratio equal to 1, charge inversion is observed ($\zeta$ potential = +19.3 ± 1.5 mV) and interactions are no longer occurring.

**Type II titration:** In Fig. 6 is presented the variation of $\zeta$ potential and binding enthalpy as a function of $TiO_2$ over alginate molar charge ratio for the titration of a 0.05 mM alginate solution with a 1 g $L^{-1}$ $TiO_2$ dispersion at pH < $pH_{PCN,TiO_2}$. Net decrease of $TiO_2$-alginate interaction takes place after six ENPs injections and, for a molar ratio equal to 1, charge inversion is observed with $\zeta$ potential values found equal to +19.3 ± 1.5 mV. Then by further increasing the $TiO_2$ concentration thermodynamic interactions are no longer observed.

A good agreement is found between the $\zeta$ potential and the interaction enthalpy values for both titration mode (type I and type II) as well as with the z-average diameter values for the type I titration. For the type I and II titrations $TiO_2$-alginate thermodynamic interactions are not occurring when charge inversion and sites saturation are observed (for Z ≥0.7 and 1, respectively) due to electrostatic repulsions and steric effects between the $TiO_2$-alginate agglomerates and the titrant. Hence simple dilution effect is observed. The main difference between the two types of titration is the molar charge ratio needed to achieve charge inversion (Z ≈ 0.5 for type I whereas Z ≈ 1 for type II). When alginate is added to the $TiO_2$ dispersion, alginate is bridging the $TiO_2$ as shown by z-average diameter values increase for Z ≤0.5 which confirms the bridging mechanism of agglomeration. Dynamic light scattering permits to assign the real-time thermogram signature occurring prior non-interaction domain to an important precipitation domain in agreement with z-average diameter and $\zeta$ potential values.

## 4. Conclusion

The association process between $TiO_2$ nanoparticles and alginate is found here mainly driven by an important gain of entropy due to the release of alginate counter-ions and water molecules. Our results also suggest that the change of binding enthalpy, via electrostatic interactions, is favorable for the binding process and that this thermodynamic value is independent of concentration and mixing order. The mixing order is nevertheless found to play a key role on the reaction stoichiometry and for the molar charge ratio needed to fully saturate the binding sites available for complexation. This is due to different mechanisms of association (charge neutralization when $TiO_2$ is added to alginate, bridging when alginate is added to $TiO_2$). ITC measurements allows to determine and explicitly quantify important thermodynamic parameters ($\Delta H_b$, $K_b$, $\Delta G$ and $\Delta S$) and to propose, when associated with light scattering techniques, different mechanisms of interactions depending on the mixing order between engineered nanoparticles and natural organic matter. It should be noted that thermodynamic properties of adsorption could also be dependent on the particle size and type as well as surface site distribution.

## Acknowledgments

The authors are grateful to the financial support received from the Swiss National Foundation (200020_152847 and 200021_135240). The work leading to these results also received funding from the European Union Seventh Framework Programme (FP7/2007-20013) under agreement no NMP4-LA-2013-310451. L.V. also thanks the CNPq (Conselho Nacional de Desenvolvimento Científico e Tecnológico) in Brazil for postdoctoral fellowship.

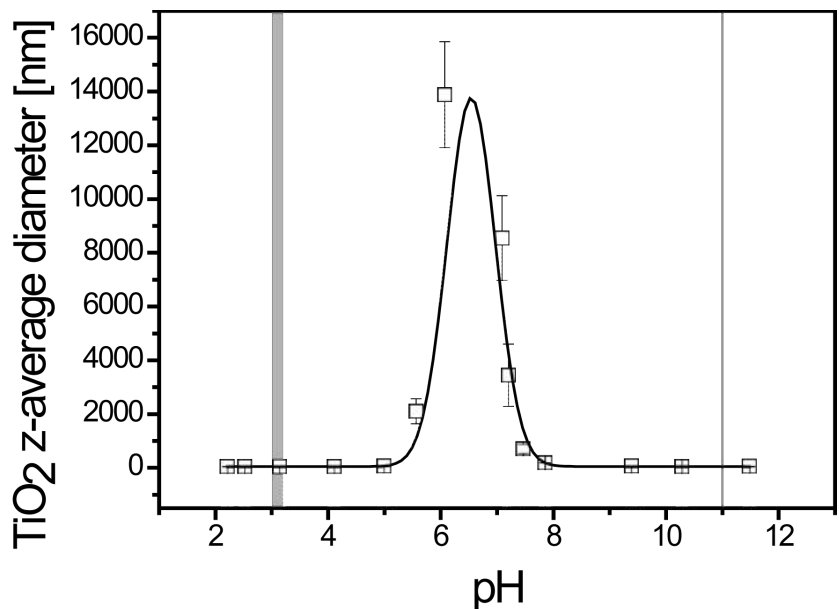

**Fig. S1** - TiO$_2$ z-average diameter values as a function of pH. At pH 3.1 (large gray vertical line) the TiO$_2$ ENPs are dispersed with a z-average diameter value found equal to 47 ± 1 nm. At pH 11.0 (narrow gray line), the ENPs are also stable with diameter value equal to 53 ± 1 nm. [TiO$_2$] = 50 mg L$^{-1}$ and [NaCl] = 0.001 M.

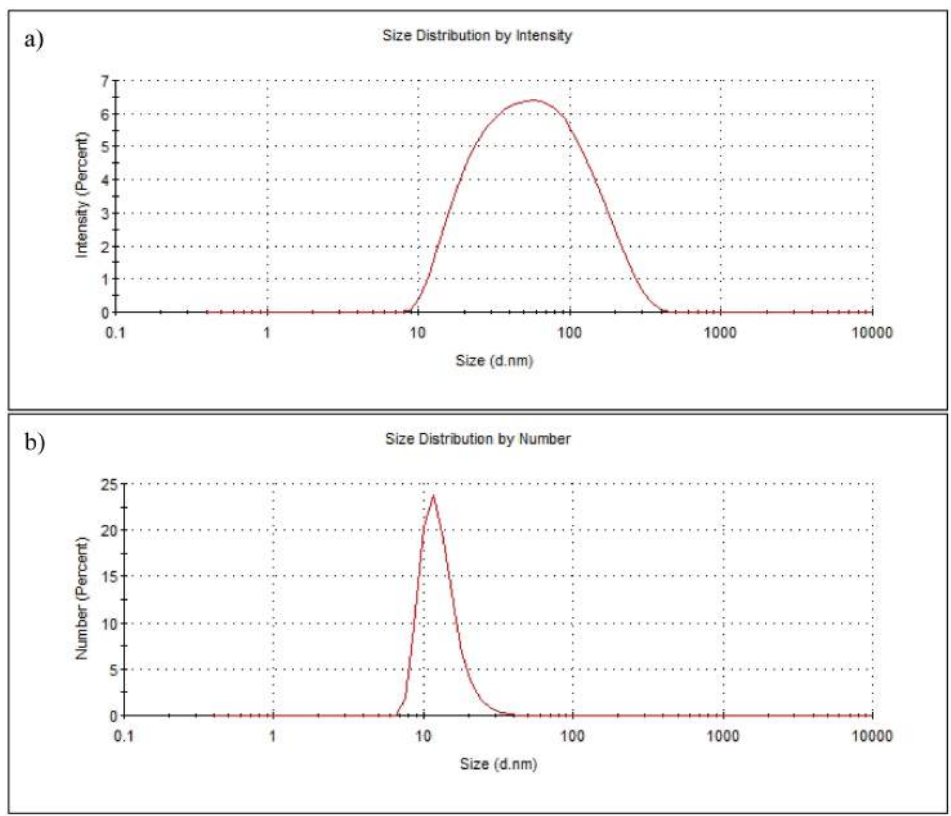

**Fig. S2 -** TiO$_2$ ENPs intensity and number size distribution at pH < pH$_{PCN,ENPs}$. [TiO$_2$] = 50 mg L$^{-1}$.

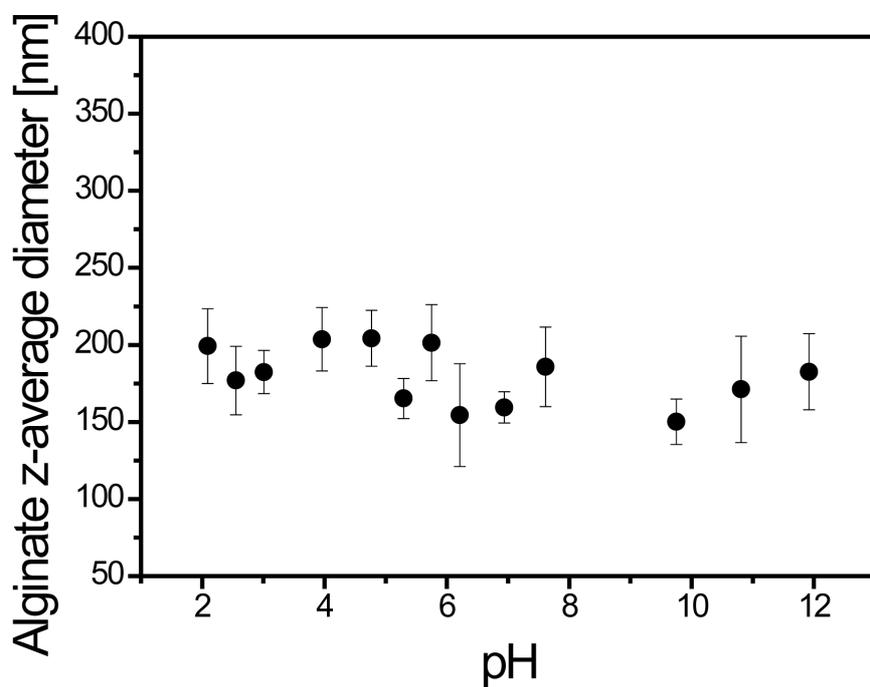

**Fig. S3** - Alginate z-average diameter values as a function of pH. Alginate z-average diameter is constant with diameter value equal to 178 ± 21 nm. [Alginate] = 100 mg L$^{-1}$.

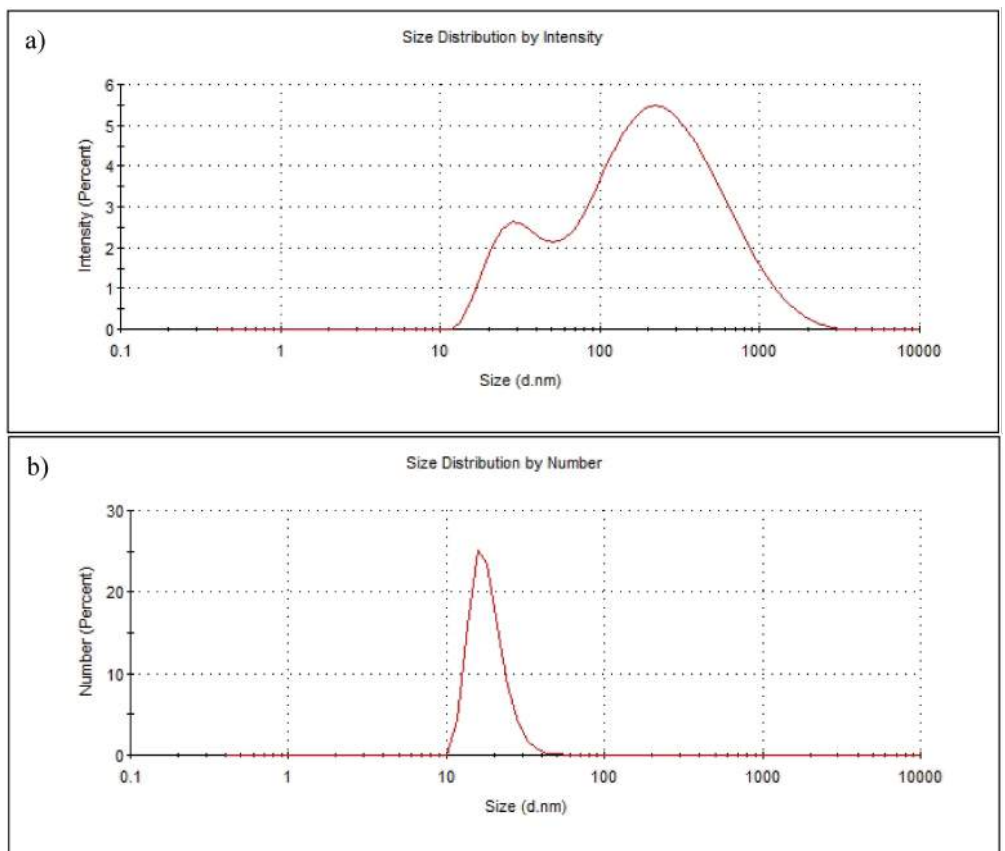

**Fig. S4 -** Alginate intensity and number size distribution at pH < $pH_{PCN,ENPs}$. [Alginate] = 100 mg $L^{-1}$.

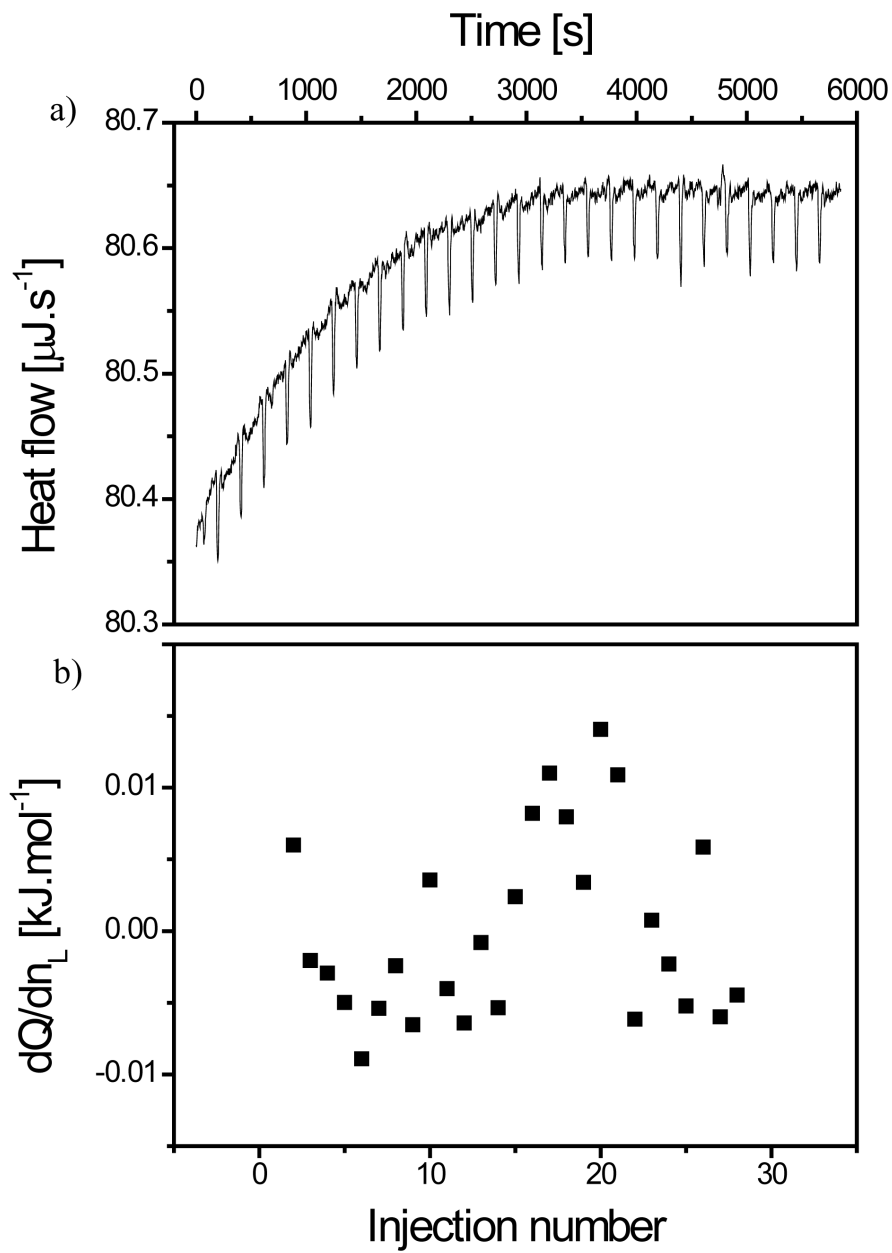

**Fig. S5** - a) Real-time thermogram representing the heat exchange for a 2.5 mM alginate charge concentration titration in water at pH < $pH_{PCN,TiO_2}$. b) Corresponding interaction binding enthalpy as a function of injection number. Only dilution effect is observed.

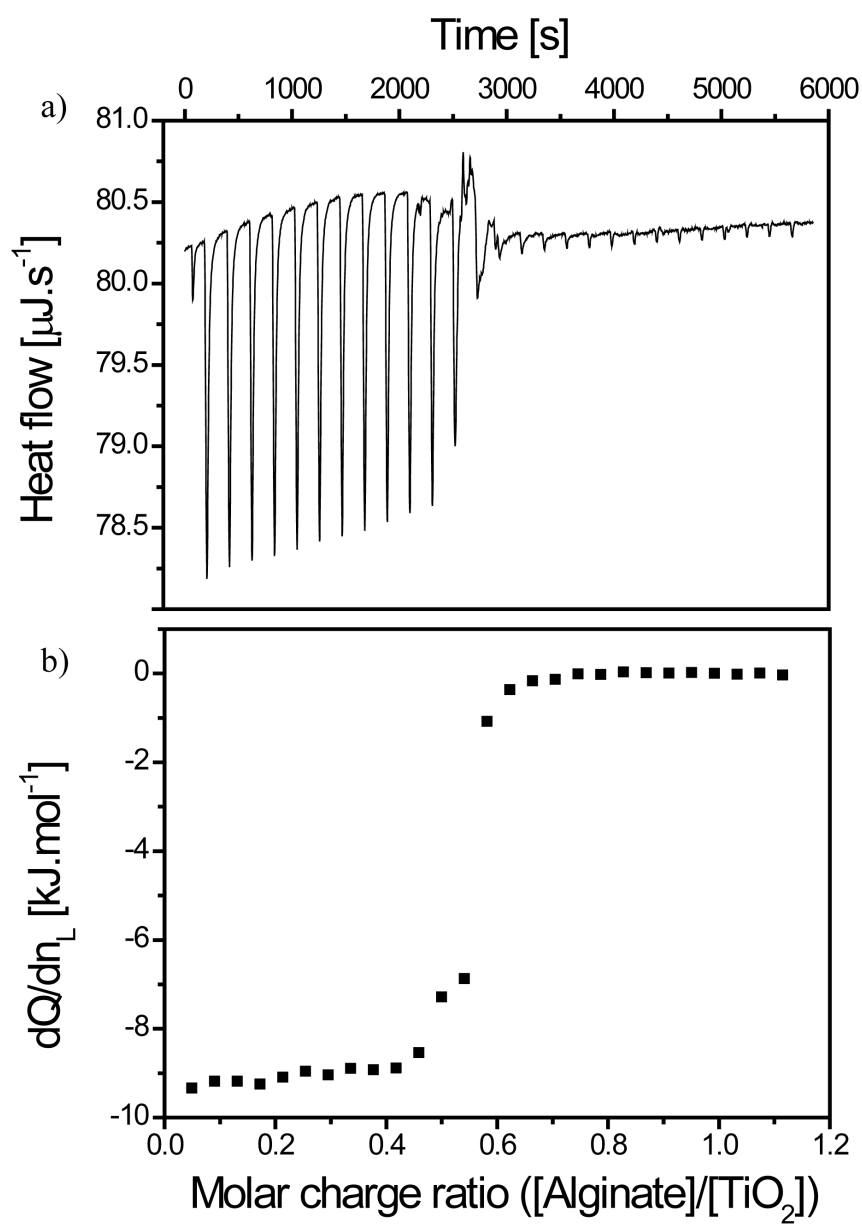

**Fig. S6 -** a) Real-time thermogram for $TiO_2$ 0.14 g $L^{-1}$ titration with alginate 0.7 mM at pH < $pH_{PCN,TiO_2}$ at 298.15 K. Negative peaks indicate an exothermic reaction. After about fifteen injections sites saturation occurs and only dilution effect is observed (small negative peaks). b) Corresponding integrated heat data as a function of molar charge ratio.

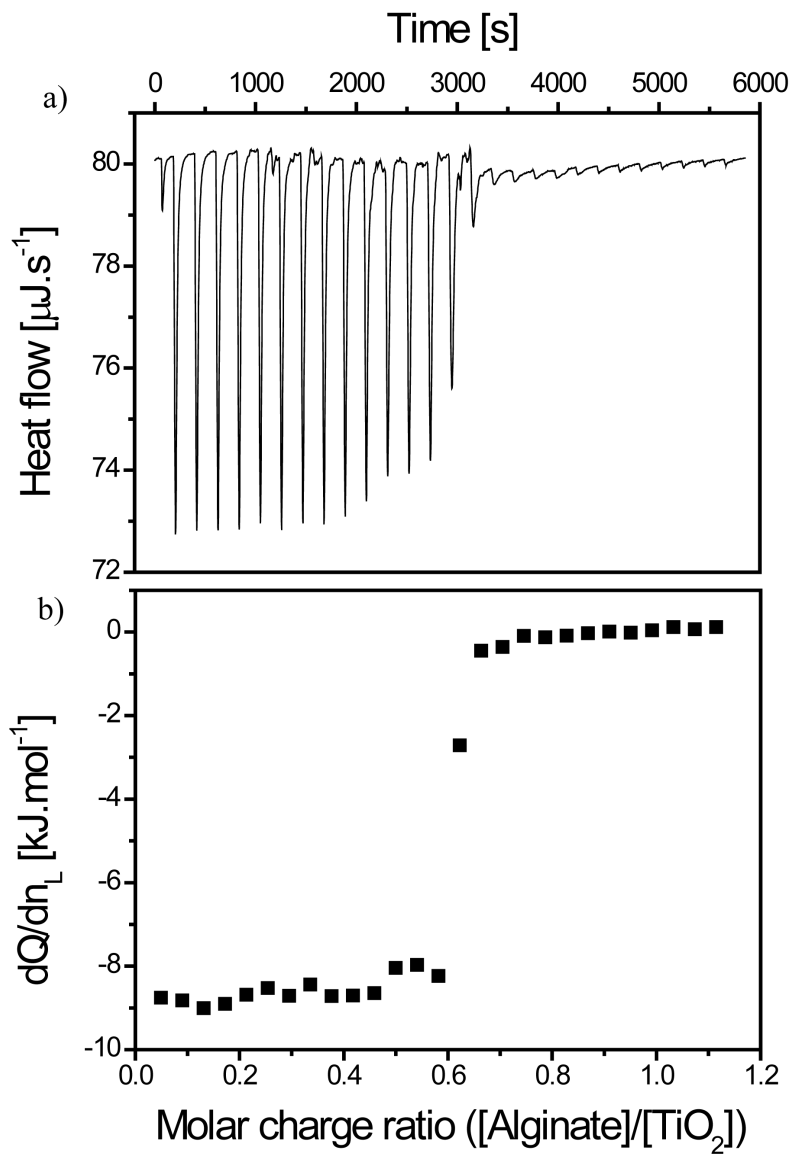

**Fig. S7** - a) Real-time thermogram for TiO$_2$ 0.5 g L$^{-1}$ titration with alginate 2.5 mM at pH < pH$_{PCN,TiO_2}$ at 298.15 K. Negative peaks indicate an exothermic reaction. After about fifteen injections sites saturation occurs and only dilution effect is observed (small negative peaks). b) Corresponding integrated heat data as a function of molar charge ratio.

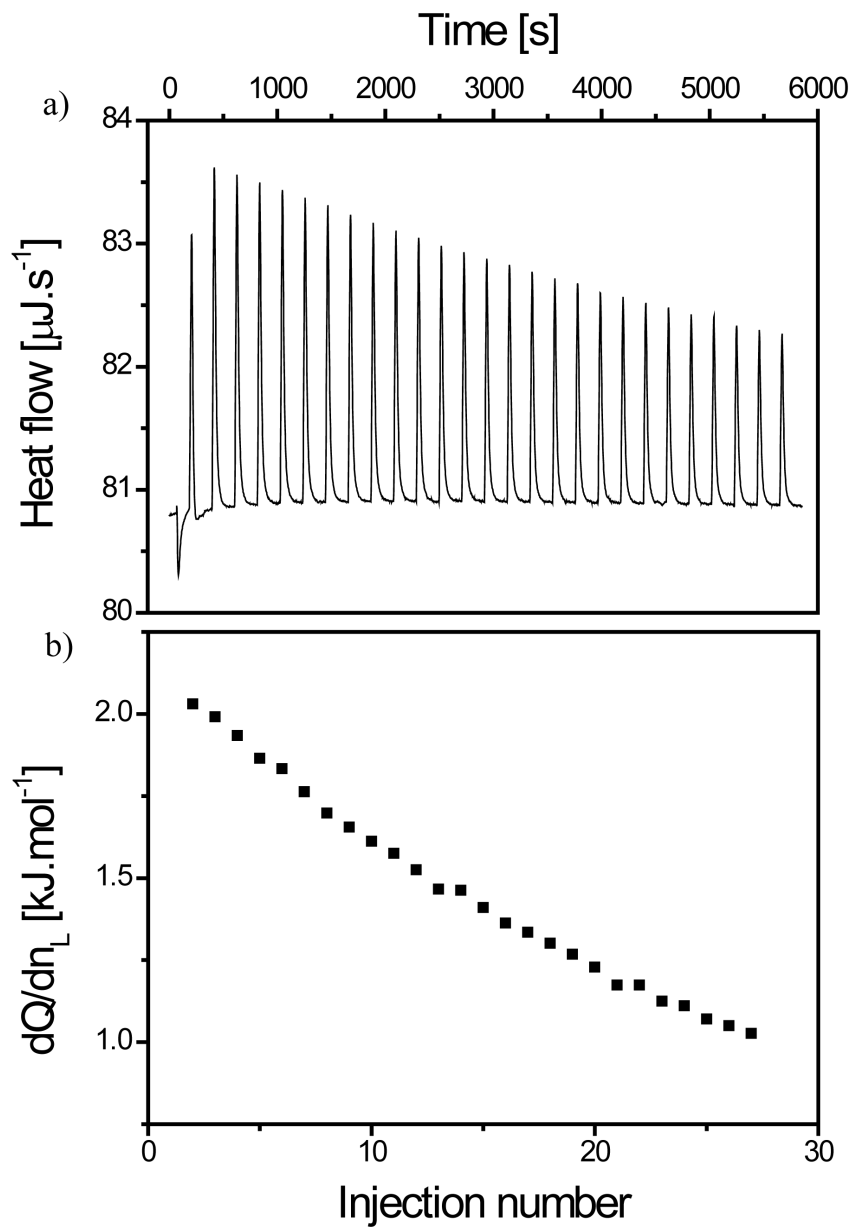

**Fig. S8** - a) Real-time thermogram and b) the corresponding integrated heat data as a function of injection number for a 5 g L$^{-1}$ TiO$_2$ dispersion titration in water at pH < pH$_{PCN,TiO_2}$ and at 298.15 K. Dilution effect is observed. The dilution of the TiO$_2$ ENPs in water is an endothermic process.

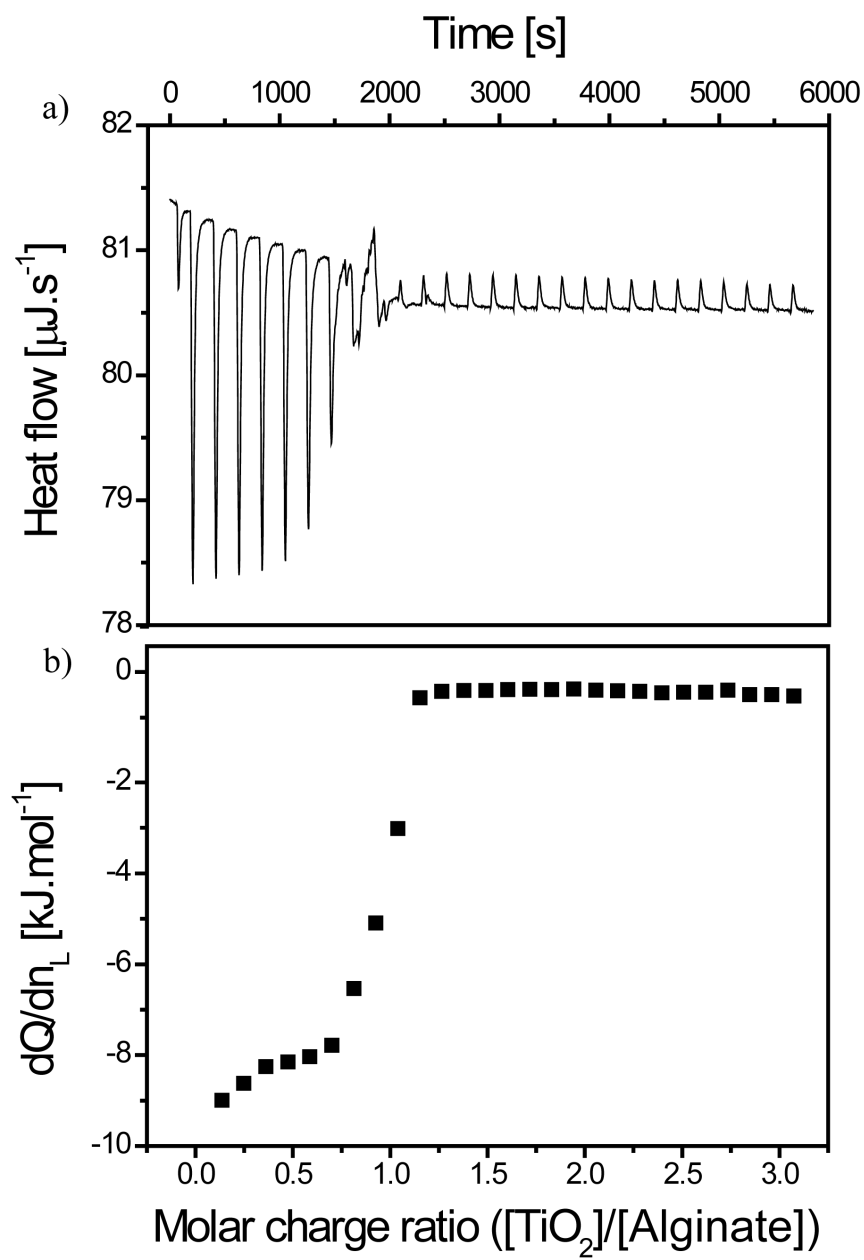

**Fig. S9** - a) Real-time thermogram for alginate 0.07 mM titration with $TiO_2$ 1.4 g $L^{-1}$ at pH < $pH_{PCN,TiO_2}$ and at 298.15 K. Negative peaks indicate an exothermic reaction. After about ten injections sites saturation occurs and only dilution effect is observed (small positive peaks). b) Corresponding integrated heat data as a function of molar charge ratio.

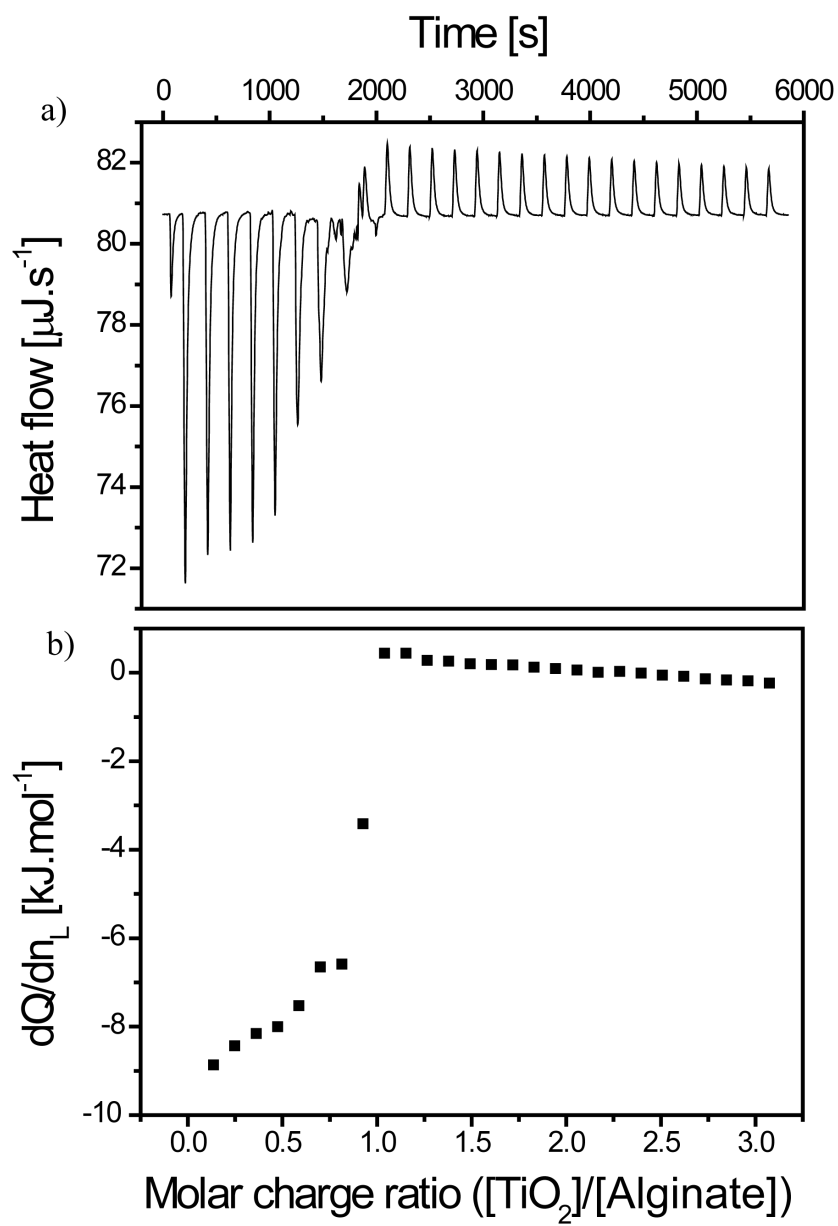

**Fig. S10** - a) Real-time thermogram for alginate 0.25 mM titration with TiO$_2$ 5 g L$^{-1}$ at pH < pH$_{PCN,TiO_2}$ and at 298.15 K. Negative peaks indicate an exothermic reaction. After about ten injections sites saturation occurs and only dilution effect is observed (small positive peaks). b) Corresponding integrated heat data as a function of molar charge ratio

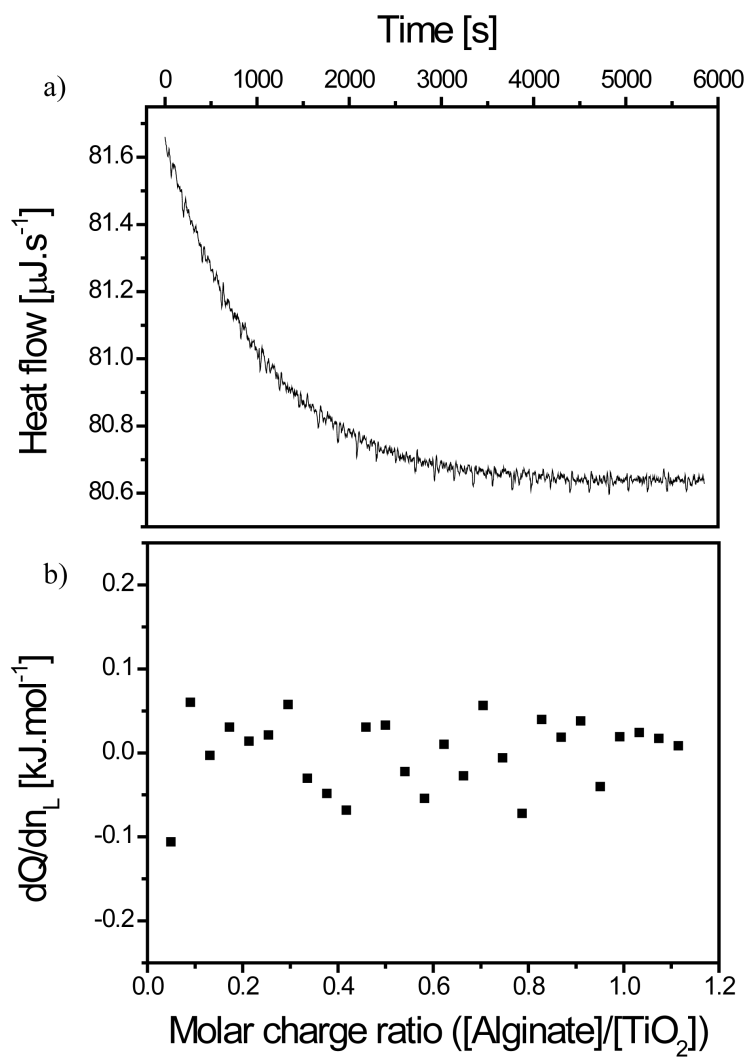

**Fig. S11** - a) Real-time thermogram representing the heat exchange for a 0.1 g L$^{-1}$ TiO$_2$ titration with a 0.5 mM charge concentration alginate at pH 11.0. b) Corresponding integrated heat exchange data as a function of alginate over TiO$_2$. No interaction is observed in agreement with electrostatic repulsions between the negatively charged compounds (Fig. 2).